\documentclass[twocolumn]{aastex701}
\usepackage{amsmath}
\usepackage{booktabs}
\usepackage{multirow}
\usepackage{color}
\usepackage{soul}
\usepackage{float}

\usepackage[dvipsnames]{xcolor} 

\defcitealias{Shen2018}{S18}

\renewcommand{\arraystretch}{1.2}  
\usepackage{listings}
\usepackage{color}
\definecolor{dkgreen}{rgb}{0,0.6,0}
\definecolor{gray}{rgb}{0.5,0.5,0.5}
\definecolor{mauve}{rgb}{0.58,0,0.82}
\definecolor{golden}{rgb}{0.86,0.65,0.01}
\usepackage{orcidlink}

\begin{document}


\title{Active galactic nuclei do not exhibit strictly sinusoidal brightness variations}

\correspondingauthor{Kareem El-Badry}
\email{kelbadry@caltech.edu}

\author[0000-0002-6871-1752]{Kareem El-Badry}
\affiliation{Department of Astronomy, California Institute of Technology, 1200 E. California Blvd., Pasadena, CA 91125, USA}
\affiliation{Max Planck Institute for Astronomy, Königstuhl 17, 69117 Heidelberg, Germany}
\email{kelbadry@caltech.edu}

\author[0000-0003-2866-9403]{David W. Hogg}
\affiliation{Max Planck Institute for Astronomy, Königstuhl 17, 69117 Heidelberg, Germany}
\affiliation{Center for Computational Astrophysics, Flatiron Institute, 160 Fifth Ave, New York, NY 10010, USA}
\affiliation{Department of Physics, New York University, 726 Broadway, New York, NY 10003, USA}
\email{david.hogg@nyu.edu}

\author[0000-0003-4996-9069]{Hans-Walter Rix}
\affiliation{Max Planck Institute for Astronomy, Königstuhl 17, 69117 Heidelberg, Germany}
\email{rix@mpia.de}

\begin{abstract}\noindent
Periodic variability in active galactic nuclei (AGN) light curves has been proposed as a signature of close supermassive black hole (SMBH) binaries.
Recently, 181 candidate SMBH binaries were identified in \textsl{Gaia} DR3 based on apparently stable sinusoidal variability in their $\sim$1000-day light curves.
By supplementing \textsl{Gaia} photometry with longer-baseline light curves from the Zwicky Transient Facility (ZTF) and the Catalina Real Time Transient Survey (CRTS), we test whether the reported periodic signals persist beyond the \textsl{Gaia} DR3 time window.
We find that in all 116 cases with available ZTF data, the \textsl{Gaia}-inferred periodic model fails to predict subsequent variability, which appears stochastic rather than periodic.
The periodic candidates thus overwhelmingly appear to be false positives; red noise contamination appears to be the primary source of false detections.
We conclude that truly periodic and sinusoidal AGN variability is exceedingly rare, with at most a few in $10^6$ AGN exhibiting it on 100 to 1000 day timescales. Models predict that the \textsl{Gaia} AGN light curve sample should contain dozens of true SMBH binaries with periods within the observational baseline, so the lack of strictly periodic light curves in the sample suggests that most short-period binary AGN do not have light curves dominated by simple sinusoidal periodicity. 

\keywords{quasars: general -- quasars: supermassive black holes —– techniques: photometric}

\end{abstract}


\section{Introduction}
\label{sec:intro}
Binary supermassive black holes (SMBHs) are a natural consequence of hierarchal assembly. As galaxies merge, their SMBHs are expected to form gravitationally bound pairs, which may shrink and eventually emit low-frequency gravitational waves detectable by pulsar timing arrays \citep[e.g.][]{Agazie2023, Reardon2023} and the Laser Interferometer Space Antenna \citep[LISA;][]{Amaro-Seoane2017}. Despite decades of theoretical work predicting the formation and evolution of binary SMBHs, identifying them observationally has proven challenging \citep[see][for recent reviews]{DeRosa2019, DOrazio2023}. 

Many previous works have reported periodic and quasi-periodic variability of AGN across the electromagnetic spectrum as possible evidence of binary SMBHs with separations of order 0.01 pc \citep[e.g.][]{Sillanpaa1988, Sillanpaa1996, Raiteri2001, Valtonen2008, Graham2015,  Liu2015, Zheng2016, Bon2016, Liu2016,  Ackermann2015, Sandrinelli2016, Caplar2017,  Severgnini2018, Chen2020, Kovacevic2020, Zhu2020, Jiang2022, Fatovic2023, Zhang2022, Luo2025, Kiehlmann2025, delaParra2025, Foustoul2025}. While early studies focused on individual objects, the proliferation of wide-field time-domain surveys has led to an explosion in the number of periodic AGN candidates in the last decade \citep[][]{Graham2015b, Charisi2016, Liu2019, Chen2024}, with several hundred candidates identified to date. One can reasonably anticipate that the Rubin observatory's Legacy Survey of Space and Time \citep[LSST;][]{Ivezic2019} will enable the identification of many more candidates \citep[e.g.][]{Davis2025, Chiesa2025}.

A fundamental challenge facing periodicity searches for binary SMBHs is that even single quasars vary stochastically over a wide range of timescales.  This variability (``red noise'') can both introducing false positives -- particularly when light curves only sample a few cycles \citep[e.g.][]{Vaughan2016, Witt2022, Robnik2024} -- and prevent the detection of real periodic variability \citep[e.g.][]{Lin2025}. State-of-the-art approaches to distinguish stochastic red-noise variability from genuine periodic signals often employ Gaussian-process models with physically motivated covariance kernels, coupled with Bayesian model comparison to quantify the evidence for periodicity. Such methods are designed to account for the correlated noise structure in quasar light curves and assess whether an apparent periodicity is statistically significant \citep[e.g.,][]{Covino2019, Covino2020, Covino2022, Zhu2020, Rigamonti2025}.

Almost all AGN variability surveys to date have used data from ground-based surveys, including the Catalina Real-Time Transient Survey \citep[CRTS;][]{Drake2009}, the Palomar Transient Factory \citep[PTF;][]{Law2009}, the Panoramic Survey Telescope and Rapid Response
System \citep[Pan-STARRS;][]{Chambers2016}, and the Zwicky Transient Facility \citep[ZTF;][]{Bellm2019}. When analyzed together, these surveys can provide light curves with time baselines exceeding 20 years and $\sim$1000 photometric measurements per object. 

The ESA \textsl{Gaia} mission \citep{GaiaCollaboration2016} produces light curves that are complimentary to ground-based surveys and, until recently, have seen limited use for studies of AGN. \textsl{Gaia} scanned the sky continuously between 2014 and 2025, with a typical cadence of one observation per month. Only the first $\sim 25\%$ of the data have been published as of the mission's 3rd data release \citep[DR3;][]{GaiaCollaboration2023, Carnerero2023}, so the currently available light curves span about 1000 days and typically have $\sim 30$ observations per source. Although the time baseline of DR3 is shorter than that of several ground-based surveys, and the typical number of measurements per source is significantly smaller, \textsl{Gaia} benefits from having higher photometric precision for faint sources than any mature wide-field ground-based survey. For a typical AGN at $G=20$, the \textsl{Gaia} per-epoch photometric precision is 0.02 mag, while the typical per-epoch precision is 0.12 mag in ZTF, and $\gtrsim 0.5$ mag in CRTS. Most AGN are faint -- e.g., there are $\sim 100$ times more quasars with $G< 20$ than with $G < 18$ \citep[e.g.][]{Storey-Fisher2024} -- so \textsl{Gaia} dramatically increases the number of AGN with high-precision light curves. 

Recently, \citet{Huijse2025} 
searched the \textsl{Gaia} DR3 $G-$band light curves for periodic variability. Beginning with a sample of 770,110 light curves for a cleaned sample of AGN candidates, they identified 48,472 light curves exhibiting period variability with a dominant period between 100 days and $1.5 T$, where $T\approx 1000$\,d is the observing baseline for each source. To reduce contamination from stochastically varying sources, \citet{Huijse2025} fit the light curves with both a purely sinusoidal model and a Damped Random Walk (DRW) model. They calculate a Bayes factor, $B_{PR}$, for each candidate, representing the ratio between the Bayesian evidence of the periodic (sinusoidal) model relative to the DRW model. They retain candidates for which $\log(B_{PR}) > 3.0423$, resulting in a final sample of 181 binary SMBH candidates. The threshold of 3.0423 was chosen because their simulations predict that 99.9\% of light curves generated from a DRW with maximum power at a period between 100 d and $1.5T$ will have $\log(B_{PR}) < 3.0423$.  

In this paper, we analyze the \textsl{Gaia} DR3 light curves of the binary SMBH candidates from \citet{Huijse2025} together with light curves from ground-based surveys, particularly ZTF. ZTF extends the total observational baseline from 1000 to $\sim 4000$ d, and its light curves were not considered by \citet{Huijse2025} when fitting periodic models and calculating Bayes factors. The ZTF light curves thus provide an opportunity to validate the models fit to the \textsl{Gaia} DR3 data.

\section{Light curves}
\label{sec:lcs}

\subsection{Gaia}
\label{sec:Gaia}

We queried the \textsl{Gaia} DR3 light curves for the 181 periodic AGN candidates from \citet{Huijse2025} using the epoch photometry Datalink service \citep{Canovas2024_Datalink}. We only considered the $G-$band data and employed the same pre-processing filters as \citet{Huijse2025}, including rejection of datapoints with $e_j > \overline{e} + 3\sigma_e$, where $\overline{e}$ and $\sigma_e$ are the mean and standard deviation of the photometric errors, $e_j$, of each source. The median number of epochs per source is 33, over a median time baseline of 932 d. 

\subsection{ZTF}
\label{sec:ZTF}

We queried the public ZTF light curves in $r-$ and $g-$bands of all 181 candidates.  We exclude data with \texttt{catflags} $\geq 0$. This yields light curves for 116 objects, or 64\% of the sample. Sources without ZTF light curves are in most cases outside the survey footprint or blended with another nearby source. 

The median number of ZTF epochs per source is 453 in the $r$ band and 316 in the $g$ band. The light curves cover a median time baseline of 2350 d, or 3650 d when combined with the \textsl{Gaia} light curves. Compared to the \textsl{Gaia} light curves, the ZTF data have higher cadence but lower photometric precision. For more straightforward comparison to the \textsl{Gaia} data, we bin the ZTF data in time bins of width $\Delta t = 20$\,d when plotting light curves, averaging the measurements within each 20 d window while weighting by inverse variance.

\begin{figure*}
    \centering
    \includegraphics[width=\textwidth]{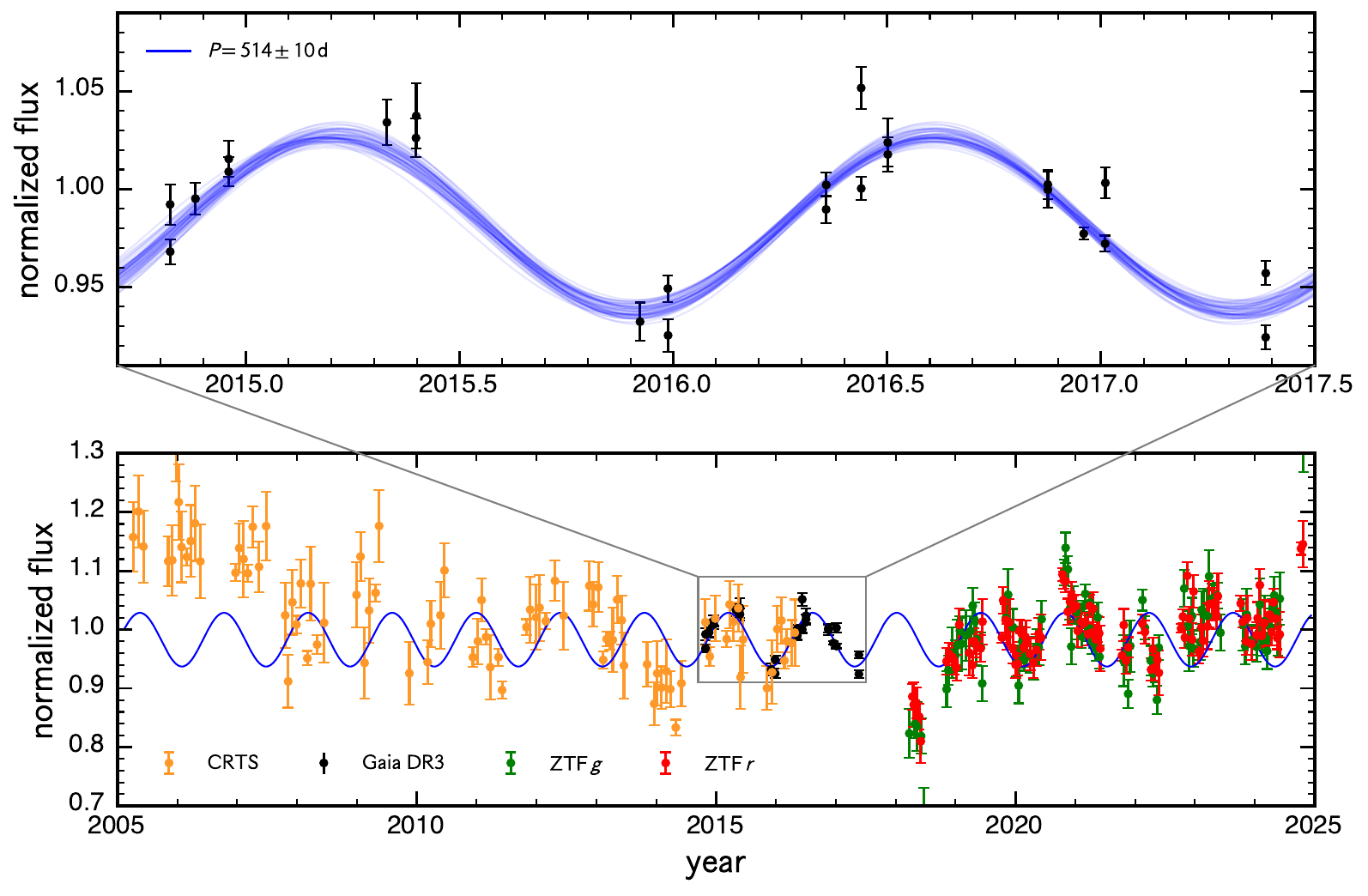}
    \caption{Long-term light curve of the source \textsl{Gaia} DR3 3870752104863883264, which \citet{Huijse2025} report as periodic with $P=520$ d and a Bayes factor $\log(B_{PR})=	3.32$, indicating strong preference of periodic variability over a damped random walk. Top panel shows the \textsl{Gaia} DR3 $G-$band light curve and posterior samples from a sinusoidal fit to it. Bottom panel combines this light curve and the best-fit sinusoid with long-term photometry from ZTF ($g-$ and $r-$ bands) and CRTS (unfiltered). Neither of these datasets is consistent with the sinusoidal model. }
    \label{fig:crts}
\end{figure*}

\subsection{CRTS}
\label{sec:CRTS}
We also queried light curves from CRTS. Most sources in the sample are too faint to have useful CRTS light curves, but three objects with $G < 19$ have well-sampled light curves spanning more than a decade.  CRTS magnitudes are unfiltered but are transformed to $V$ using an empirically calibrated, color-dependent relation. As with the ZTF data, we binned the light curves in 20-day windows. 

\section{Analysis}
\label{sec:analysis}

In Figure~\ref{fig:crts}, we show the light curve of \textsl{Gaia} DR3 source ID 3870752104863883264. At $G=18.4$, this is one of the few sources in the sample bright enough to have a useful CRTS light curve. The top panel shows the \textsl{Gaia} DR3 data and posterior samples from fitting it with a sinusoidal model. The fit is carried out with \texttt{emcee} \citep{Foreman-Mackey2013}, with the period, variability amplitude, phase, and mean value left free. The inferred period is $P=514\pm 10$\,d, which is consistent with the best-fit period of 520\,d reported by \citet{Huijse2025} for this source. Over the \textsl{Gaia} DR3 time baseline, the light curve indeed resembles a sinusoid, although the presence of a few datapoints that are offset from the model predictions by two or three times their uncertainties suggests the presence of additional variability.

In the bottom panel of Figure~\ref{fig:crts}, we show the light curves from CRTS and ZTF as well. The blue curve extends the best-fit sinusoid fit to the \textsl{Gaia} data alone. The CRTS light curve overlaps with most of the \textsl{Gaia} observation window and includes variability consistent with the sinusoidal model during that period. However, the agreement ceases immediately before the DR3 observation window. ZTF observations begin about a year after the DR3 window and are again immediately inconsistent with the sinusoidal model. The variability in the $r-$ and $g-$ bands is very similar. 

The upshot is that the source shown in Figure~\ref{fig:crts} does not exhibit stable sinusoidal variability, but rather varies stochastically. The existence of one such source in the sample is not surprising, since even optimistic interpretation of the simulations carried out by \citet{Huijse2025} would predict that the sample should contain $48,472\times 0.001\approx 48$ false positives. However, inspection of the light curves for other sources suggests that false positives actually dominate the sample. 

Figure~\ref{fig:ztf} shows the \textsl{Gaia} and ZTF light curves of all 116 sources with ZTF data. In essentially all cases, the ZTF data are manifestly inconsistent with extrapolation of the best-fit sinusoidal model. For most sources, the disagreement begins immediately, implying that the sinusoidal model is not predictive and \textsl{Gaia} DR4 data will also be inconsistent with the best-fit sinusoidal model. 
While the \textsl{Gaia}  light curves generally have much higher per-epoch precision -- particularly for the faint sources in the later panels --  ZTF provides denser temporal sampling and a longer baseline. A \textsl{Gaia}  light curve of a stably periodic AGN, if observed with ZTF's cadence and photometric uncertainties, would still exhibit clearly detectable periodicity at the amplitudes inferred from \textsl{Gaia} ; the fact that no such coherence is seen therefore reflects an intrinsic absence of stable periodicity rather than differences in data quality.

\begin{figure}
    \centering
    \includegraphics[width=0.5\textwidth]{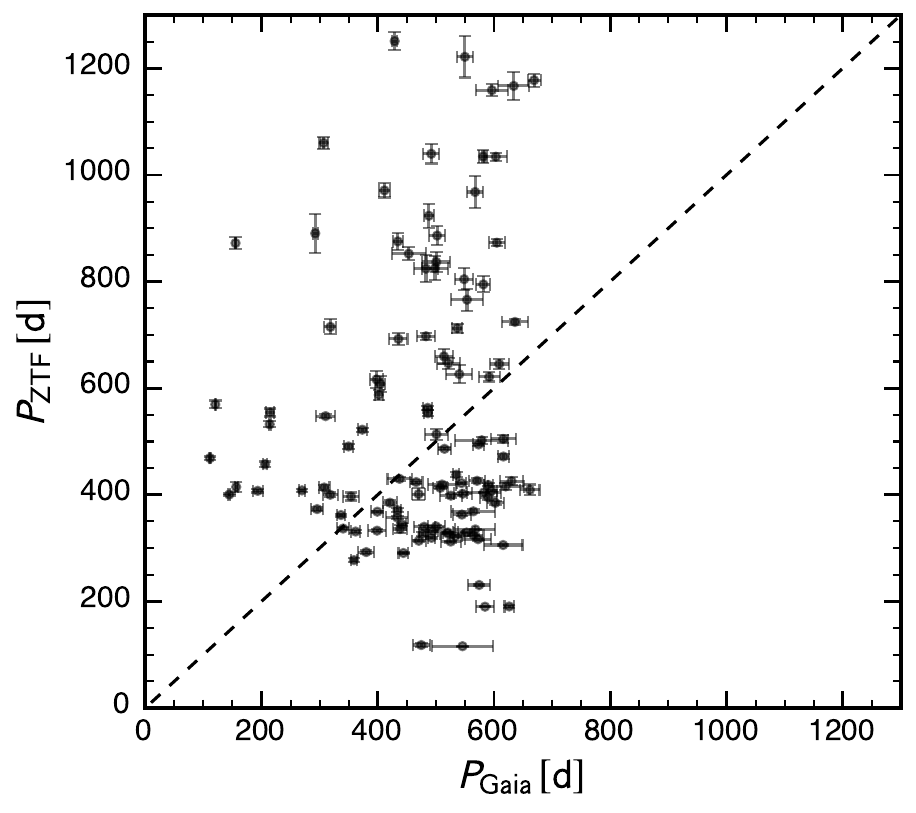}
    \caption{Best-fit periods measured from \textsl{Gaia} and ZTF light curves for all 116 sources with ZTF data. Errorbars show the formal uncertainties when fitting light curves with a sinusoid. The cutoff at $P_{\rm Gaia} \lesssim 700$\,d is a consequence of cuts on the input sample. There is no correlation between the \textsl{Gaia} period and the dominant period measured in the ZTF light curves for the same objects five years later. Three (six) sources have best-fit periods that are consistent across the two surveys within one (two) sigma.   }
    \label{fig:periods}
\end{figure}

\subsection{Consistency of the periods}
To assess the consistency of periodicity between the ZTF and \textsl{Gaia} data, we determine a best-fit period for each unbinned ZTF light curve and compare it to the \textsl{Gaia} value. Because the ZTF variability is nearly achromatic, we normalize and concatenate the $g$- and $r$-band  data into a single normalized light curve for each source. This reduces noise for sources exhibiting genuine stable periodicity makes possible a one-to-one comparison between ZTF and \textsl{Gaia}  periods but does not significantly change our conclusions about the number of sources with consistent periods between the two datasets. We compute a Lomb–Scargle periodogram \citep{Lomb1976, Scargle1982} on a uniform frequency grid spanning $1/1000$ to $1/100\,{\rm d^{-1}}$ and adopt the period of maximum power as the candidate ZTF period. We refine this period and estimate its uncertainty by fitting the data with a sinusoidal model that includes a ``scatter'' term for each source which is added to the photometric uncertainties in quadrature. We initialize the period of the sinusoid at the period of peak periodogram power but allow it to vary as a free parameter. This allows the best-fit periods of a few sources to be larger than the 1000-d maximum period on which we evaluate the periodogram.

For more than 90\% of sources in the sample, the peak power corresponds to a false alarm probability (FAP) $<10^{-3}$, whether the FAP is estimated using the analytic expression of \citet{Baluev2008} for Gaussian white noise or via bootstrap resampling of the light curves. Nevertheless, most of the ZTF light curves do not appear convincingly periodic (Figure~\ref{fig:ztf}), and we emphasize that both the analytic and bootstrap FAP estimates can be unreliable in the presence of red noise. 

Figure~\ref{fig:periods} compares the thus-measured ZTF periods to the periods estimated from the same sources' \textsl{Gaia} light curves. Only three objects have periods between the two datasets that are consistent within $1\sigma$; they have source IDs 3611046942987407232, 44567982277833984, and 6802710021946319488. Another three objects have periods consistent within $2\sigma$: source IDs 3995957307849856640, 723111620351503872, and 1543792150205554688. Inspection of these sources' light curves (Figure~\ref{fig:ztf}) shows that although both the \textsl{Gaia} and ZTF light curves have power at the same periods, they are not consistent with being drawn from a single sinusoid without additional red noise. The most promising of these sources is 1543792150205554688, a faint source with $G=20.6$ and relatively large ZTF photometric uncertainties, but its light curve deviates systematically from the \textsl{Gaia} prediction in the last year of observations. 

If we randomly shuffle the \textsl{Gaia} and ZTF periods, we find that $5\pm 2$ and $9\pm 3$ sources have periods consistent within one and two sigma, respectively. Finding three and six sources with consistent periods in the sample is thus entirely consistent with the periods being completely independent. We also checked whether, for sources with a dominant ZTF period different from the \textsl{Gaia}  period, a less significant peak can be found in the ZTF periodgram at the \textsl{Gaia}  period. While this does occasionally occur, we find that the typical power at the \textsl{Gaia}  period in the ZTF periodogram is similar to the typical power if we randomly shuffle the \textsl{Gaia}  periods and repeat the exercise. In particular, only $55\pm 5$ out of 116 sources have more power at the \textsl{Gaia}  period than at the \textsl{Gaia}  period of a randomly selected different source in the sample. These results imply that power at the \textsl{Gaia}  periods is consistent with noise. 

\section{Summary and Discussion}
We have shown that a large majority of periodic AGN candidates from \textsl{Gaia} light curves do not exhibit stably sinusoidal variability on timescales longer than the 1000-d baseline of the data from which they were selected. \citet{Huijse2025} argue that their approach of selecting candidates well-fit by an inflexible periodic model without red noise leads to a ``purity-driven'' sample with minimal false positives, at the possible expense of under-selection of true candidates. The failure of the periodic model to predict future data shows that the sample nevertheless has low purity. This suggests that the DRW model that \citet{Huijse2025} used in simulations under-predicts the incidence of apparent periodicity on short timescales. 

These findings highlight the importance of extended time baselines and multi-survey validation in searches for periodic AGN. Periodic signals inferred from a few cycles of data -- particularly in sparsely sampled light curves -- are highly vulnerable to contamination from stochastic variability. Even with a conservative threshold on Bayesian model comparison, we find that the vast majority of candidates selected from \textsl{Gaia} DR3 fail to exhibit coherent variability when additional data from ZTF are incorporated. This suggests that commonly used stochastic models like the damped random walk underestimate the false positive rate. More flexible stochastic models, such as those incorporating power-law power spectra or quasi-periodic oscillations, may improve purity.

\subsection{Limits on strictly sinusoidal variability}
\textsl{Gaia} DR3 contains light curves for $N_{\rm AGN} =  7.7\times 10^5$ sources passing the cuts to be included in the initial candidate sample of \citet{Huijse2025}. Assuming that \citet{Huijse2025} selected the 181 most stably sinusoidal candidates for short-period binary SMBHs -- and our analysis implies that at most $\sim 1\%$ of these candidates (i.e., $\lesssim 1/116$, given that the sources with ZTF data are randomly selected from the parent sample) display stable sinusoidal variability without additional red noise -- we can conclude that the fraction of all AGN with stably sinusoidal variability on periods of 100-1000\,d and amplitudes greater than a few percent is at most a few $\times 10^{-6}$. Given that the \textsl{Gaia} AGN sample is $\sim 10$ times larger than any previous sample of AGN with light curves of similar quality, this upper limit is $\sim 10$ times lower than any that could be set previously. 


\subsubsection{How many binary SMBHs are in the {\slshape Gaia} sample?}

While we have shown that the \textsl{Gaia} candidates do not have strictly sinusoidal light curves, we have not ruled out the possibility that they could contain genuine binary AGN. We can conservatively estimate the expected number of binary SMBHs in the \textsl{Gaia} sample as follows. We assume AGN-hosting galaxies at $1 \lesssim z \lesssim 3$ (a range that includes most of the \textsl{Gaia} sample) experience a merger with stellar mass ratio $\gtrsim 0.1$ at a per-galaxy rate of $\mathcal{R}_{\rm merger} \approx 0.1\,{\rm Gyr^{-1}}$ \citep[e.g.][]{O'Leary2021}. After a galaxy merger, the BHs approach each other through a variety of processes, but by the time they reach $P_{\rm orb}\lesssim 600\,{\rm d}$ (as required to be included in the \citealt{Huijse2025} sample), the inspiral is dominated by gravitational waves. For a circular orbit, the inspiral time \citep{Peters1964} is: 
\begin{equation}
    \label{eq:insp}
    \tau_{\rm inspiral}=5.6\times10^{4}\frac{\left(1+q\right)^{1/3}}{q}\left(\frac{M_{{\rm BH}}}{10^{8}\,M_{\odot}}\right)^{-5/3}\left(\frac{P_{{\rm orb}}}{600\,{\rm d}}\right)^{8/3}\,{\rm yr},
\end{equation}
where $M_{\rm BH,1}$ is the mass of the more massive black hole and $q=M_{\rm BH,2}/M_{\rm BH,1}$ is the mass ratio. For a $10^8\,M_{\odot}$ primary  at $P_{\rm orb} = 600$\,d, the inspiral time ranges from $7\times 10^4$ yr for $q=1$ to $6\times 10^5$ yr for $q=0.1$. 

If we assume that every merger between massive galaxies ultimately leads to a merger of their BHs, we can estimate the number of binary SMBHs in the \textsl{Gaia} sample as

\begin{equation}
    \label{eq:Nbinarysmbh}
    N_{{\rm binary\,SMBH}}=N_{{\rm AGN}}\times \mathcal{R}_{\rm merger}\times \tau_{\rm inspiral}.
\end{equation}

Here it is implicitly assumed that $\mathcal{R}_{\rm merger}$ represents the rate of mergers that are completed and eventually end up in the gravitational wave-dominated phase.
$N_{{\rm binary\,SMBH}}$ is expected to be a function of $P_{\rm orb}$, $q$, and $M_{\rm BH,1}$, since all three terms in Equation~\ref{eq:Nbinarysmbh} depend on these quantities. The steep dependence of Equation~\ref{eq:insp} on orbital period implies that the period distribution of any observed sample of periodic binary AGN will be dominated by objects with the longest observable periods. It will also be dominated by the lowest mass ratios that lead to a detectable signal, since low mass ratio mergers are more common and lead to slower inspirals. If we adopt the $\tau_{\rm inspiral}$ predicted for $q=0.2$ as an average value for $0.1 < q < 1$, Equation~\ref{eq:Nbinarysmbh} predicts 23 binary SMBHs with $q\gtrsim 0.1$ and $P_{\rm orb} \lesssim 600\,{\rm d}$ in the \textsl{Gaia} sample. This is a conservative estimate at large $q$, because it neglects the fact that AGN are probably more likely than typical galaxies to have experienced a recent merger \citep[e.g.][]{Hopkins2008}. It is more uncertain at small $q$, because not all galaxy mergers necessarily lead to a binary SMBH.

Alternatively, we can anchor our estimate of the number of binary SMBHs in the \textsl{Gaia} sample to existing models of SMBH mergers. Using a semi-analytic model built on halo merger trees,  \citet{Sesana2008} estimate that the all-sky rate of SMBH mergers at $1 < z < 3$ with chirp mass $\mathcal{M} > 10^{7.5}\,M_{\odot}$ (corresponding to $q > 0.15$ for $M_{\rm BH,1}=10^8\,M_{\odot}$) is $\approx 0.03\,{\rm yr^{-1}}$. For each merger, the time that the binary spends with $P_{\rm orb} < 600\,{\rm d}$ can be calculated from Equation~\ref{eq:insp}. If we again adopt $M_{\rm BH,1} = 10^8\,M_{\odot}$ and $q=0.2$ as typical values, we can estimate that the total number of binary SMBHs with $P_{\rm orb} < 600\,{\rm d}$  at $1 < z < 3$ is $\approx 9000$. Not all of these will be in the \textsl{Gaia} sample, because (a) SMBHs are not always active and (b) the \textsl{Gaia} periodic AGN sample is not complete. Adopting an average AGN duty cycle of 0.05 at $1 < z < 3$ \citep{Shankar2009} and assuming 10\% of all AGN with $M_{\rm BH,1} \gtrsim 10^8\,M_{\odot}$ at $1 < z < 3$ have light curves published in the \textsl{Gaia} variable AGN sample, we finally estimate that the sample should contain 45 binary SMBHs with $0.1 \lesssim q \lesssim  1$ and $P_{\rm orb} \lesssim 600\,{\rm d}$.

These are rough estimates and depend somewhat on the detailed SMBH mass and redshift distribution of the \textsl{Gaia} sample, which we have not attempted to model in detail. The predicted number of binary SMBHs in the sample -- as estimated either from per-galaxy merger rates or from a model of the cosmic SMBH binary merger rate -- is 23 or 45. These estimates are $\gg 1$, and so despite their significant uncertainties, we conclude that the \textsl{Gaia} light curve sample is  likely to contain at least a handful of binary SMBHs with $P_{\rm orb} \lesssim 600\,{\rm d}$. 

The fact that no objects in the sample have strictly periodic, sinusoidal light curves then implies that short-period binary SMBHs do not typically have strictly periodic, sinsusoidal light curves, at least not at the amplitudes probed by \textsl{Gaia}. This is not necessarily surprising: most AGN vary stochastically, and processes that the turbulence which gives rise to this variability will often still be present in the light curves of AGN containing binary SMBHs. Taken together, these findings imply that the search for binary SMBHs will require moving beyond a paradigm of sinusoidal variability and toward methods that embrace the complex, stochastic nature of AGN light curves \citep[e.g.][]{Zhu2020}.

\section*{acknowledgments}
We thank Pablo Huijse Heise, Conny Aerts, Jordy Davelaar, Joris De Ridder, Daniel J. D'Orazio, Zoltan Haiman, and Anthony Readhead for valuable discussions, and two anonymous referees for constructive feedback. This research was supported by NSF grant AST-2307232 and a Sloan Research Fellowship in Physics.

This work presents results from the European Space Agency (ESA) space mission \textsl{Gaia} . \textsl{Gaia}  data are being processed by the \textsl{Gaia}  Data Processing and Analysis Consortium (DPAC). Funding for the DPAC is provided by national institutions, in particular the institutions participating in the \textsl{Gaia}  MultiLateral Agreement (MLA). The \textsl{Gaia}  mission website is https://www.cosmos.esa.int/gaia. The \textsl{Gaia}  archive website is https://archives.esac.esa.int/gaia.

Based on observations obtained with the Samuel Oschin 48-inch Telescope at the Palomar Observatory as part of the Zwicky Transient Facility project. ZTF is supported by the National Science Foundation under Grant No. AST-1440341 and a collaboration including Caltech, IPAC, the Weizmann Institute for Science, the Oskar Klein Center at Stockholm University, the University of Maryland, the University of Washington, Deutsches Elektronen-Synchrotron and Humboldt University, Los Alamos National Laboratories, the TANGO Consortium of Taiwan, the University of Wisconsin at Milwaukee, and Lawrence Berkeley National Laboratories. Operations are conducted by COO, IPAC, and UW.

The CSS survey is funded by the National Aeronautics and Space Administration under Grant No. NNG05GF22G issued through the Science Mission Directorate Near-Earth Objects Observations Program.  The CRTS survey is supported by the U.S.~National Science Foundation under grants AST-0909182.

The Flatiron Institute is a division of the Simons Foundation.



\bibliographystyle{aasjournal}


\appendix

\twocolumngrid


\section{All the light curves}
\label{sec:appendix}
\begin{figure}[H]
\figurenum{3}
\centering
\includegraphics[width=\columnwidth]{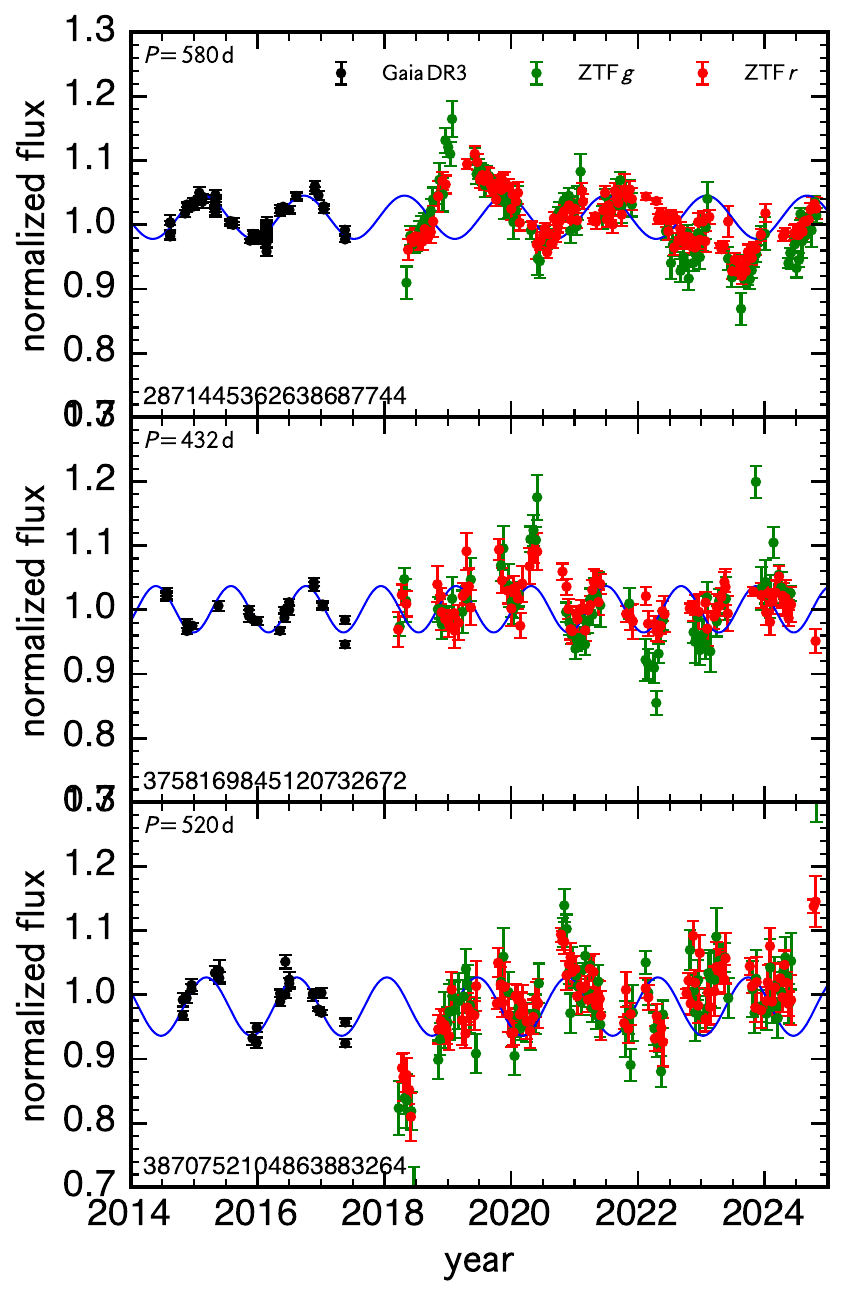}
\caption{Light curves of 116 sources. Black points show \textsl{Gaia} DR3; red and green points show binned ZTF $r-$ and $g-$band data. Blue line shows the best-fit sinusoid, with the period fixed to the value inferred by \citet{Huijse2025}. That period and the \textsl{Gaia} DR3 source ID are listed in each panel. In all cases, the sinsuidal model fails to predict the ZTF data.}
\label{fig:ztf}
\end{figure}

Figure~\ref{fig:ztf} shows \textsl{Gaia}  and ZTF light curves of all the 116 sources in our sample. Sources are sorted by increasing $G-$band magnitude.

\begin{figure}[H]
\figurenum{3}
\centering
\includegraphics[width=\columnwidth]{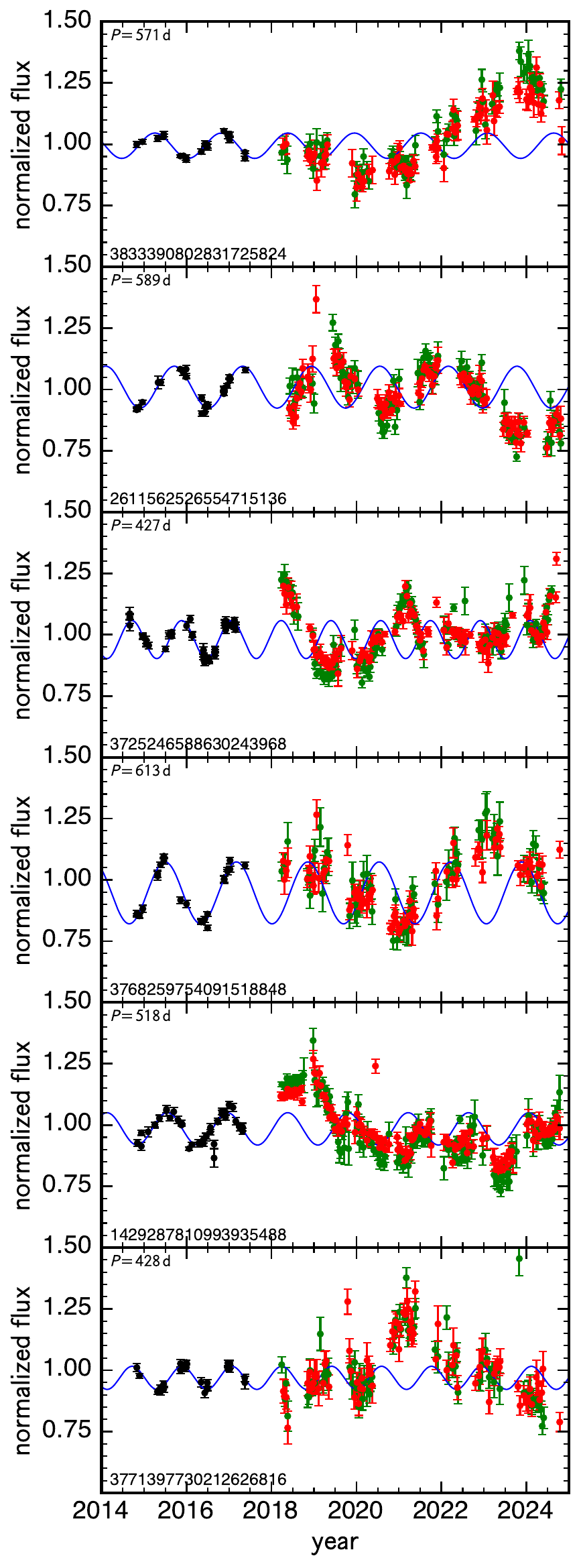}
\caption{Figure 3 (continued)}
\end{figure}

\begin{figure}[H]
\figurenum{3}
\centering
\includegraphics[width=\columnwidth]{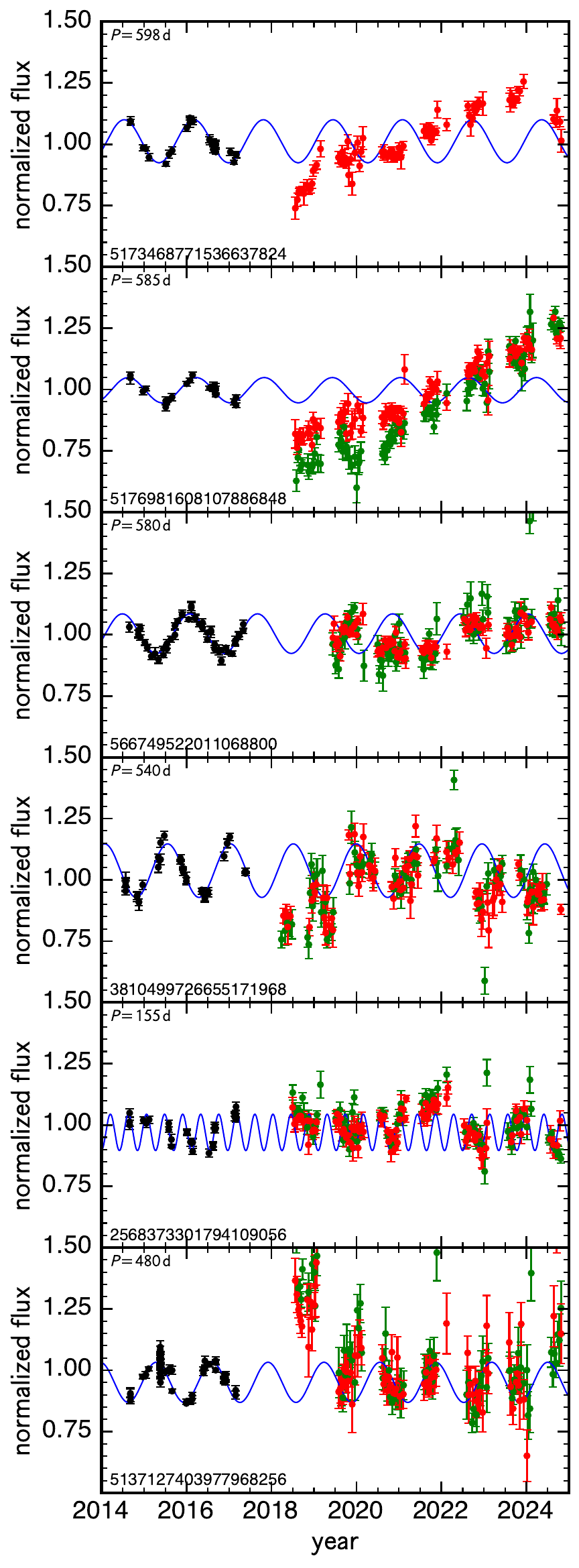}
\caption{Figure 3 (continued)}
\end{figure}

\begin{figure}[H]
\figurenum{3}
\centering
\includegraphics[width=\columnwidth]{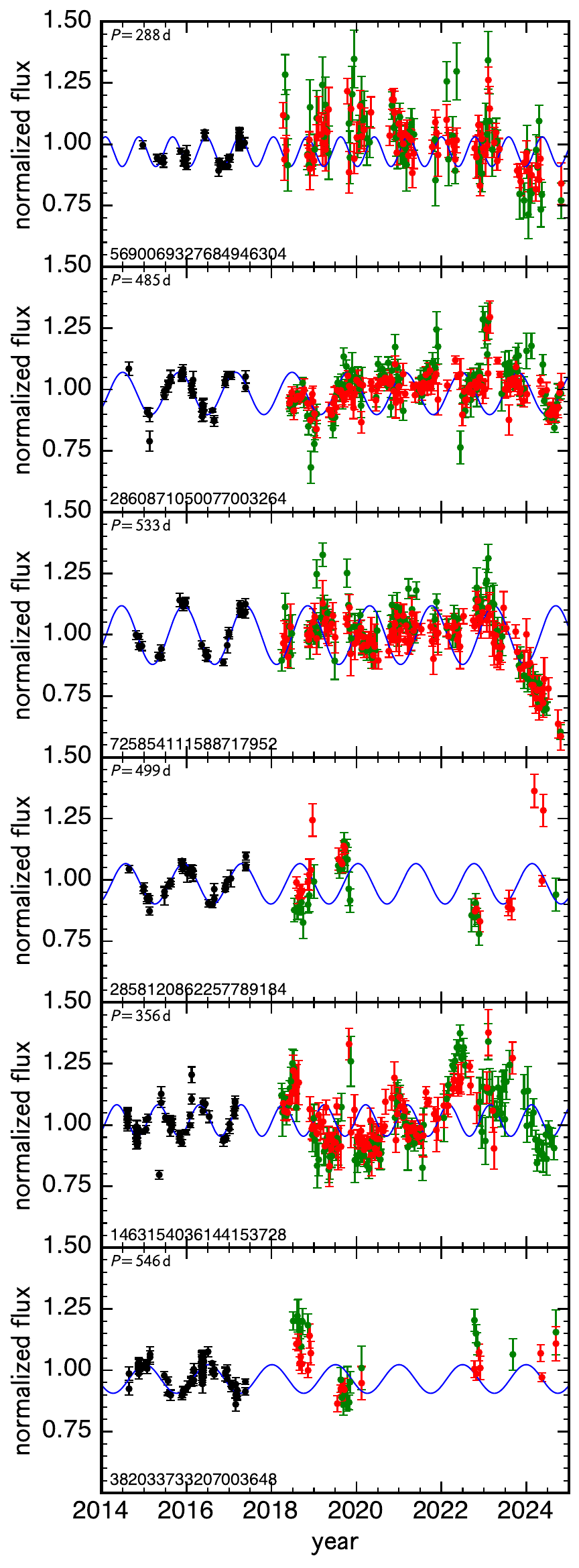}
\caption{Figure 3 (continued)}
\end{figure}

\begin{figure}[H]
\figurenum{3}
\centering
\includegraphics[width=\columnwidth]{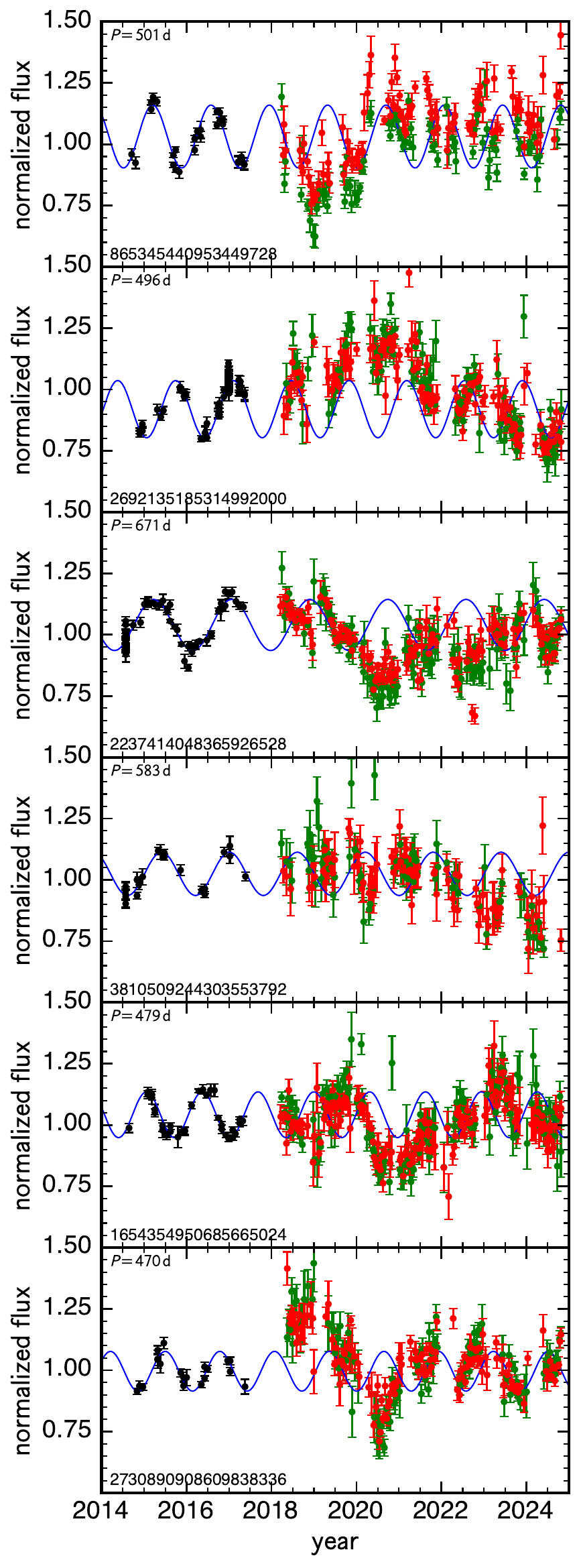}
\caption{Figure 3 (continued)}
\end{figure}

\begin{figure}[H]
\figurenum{3}
\centering
\includegraphics[width=\columnwidth]{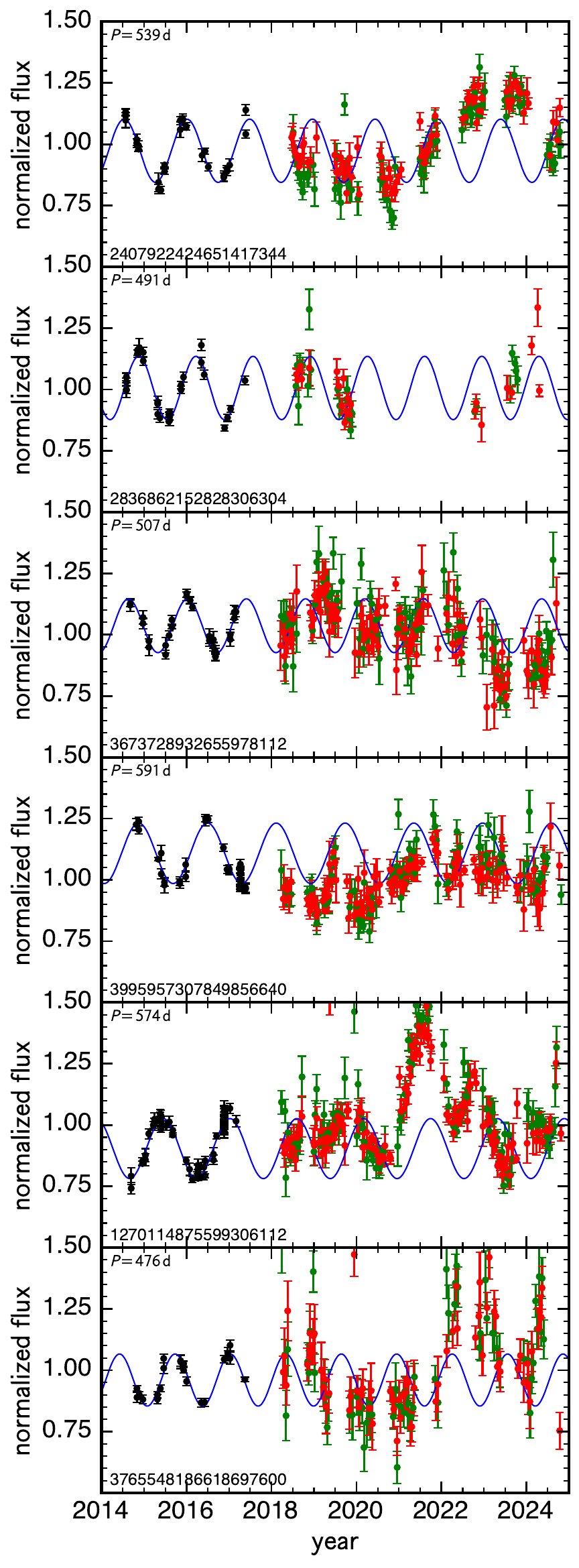}
\caption{Figure 3 (continued)}
\end{figure}

\begin{figure}[H]
\figurenum{3}
\centering
\includegraphics[width=\columnwidth]{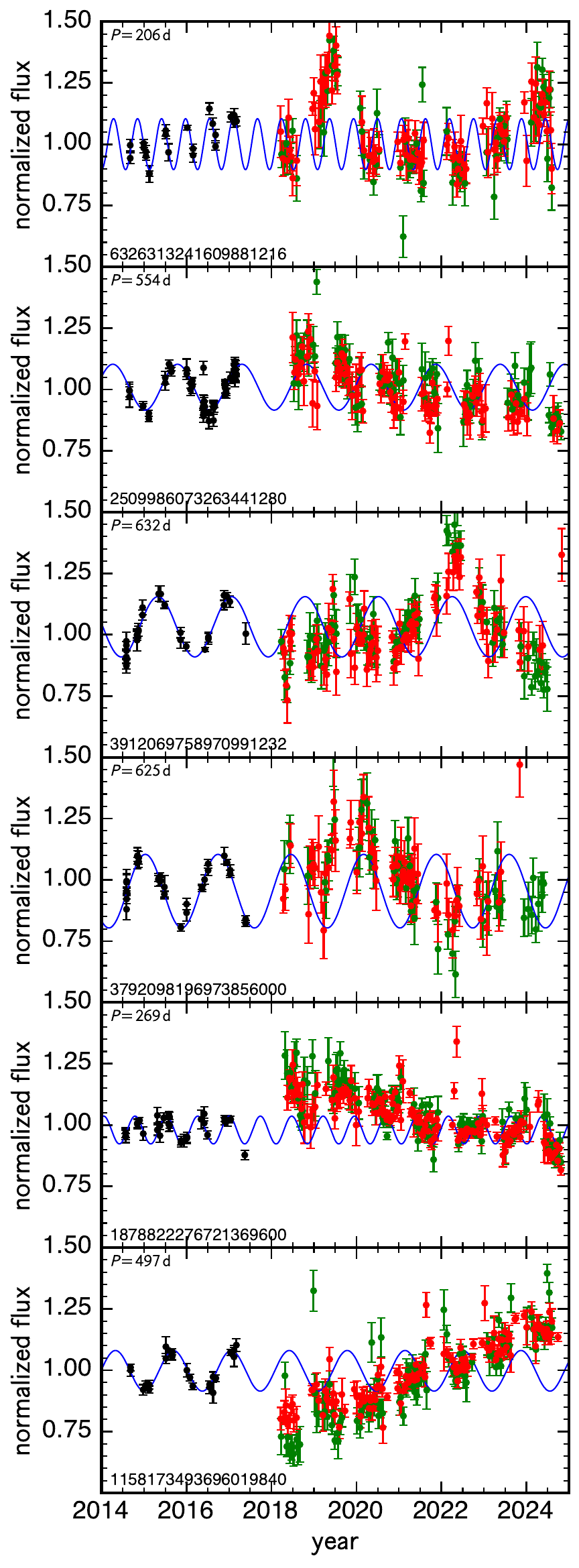}
\caption{Figure 3 (continued)}
\end{figure}

\begin{figure}[H]
\figurenum{3}
\centering
\includegraphics[width=\columnwidth]{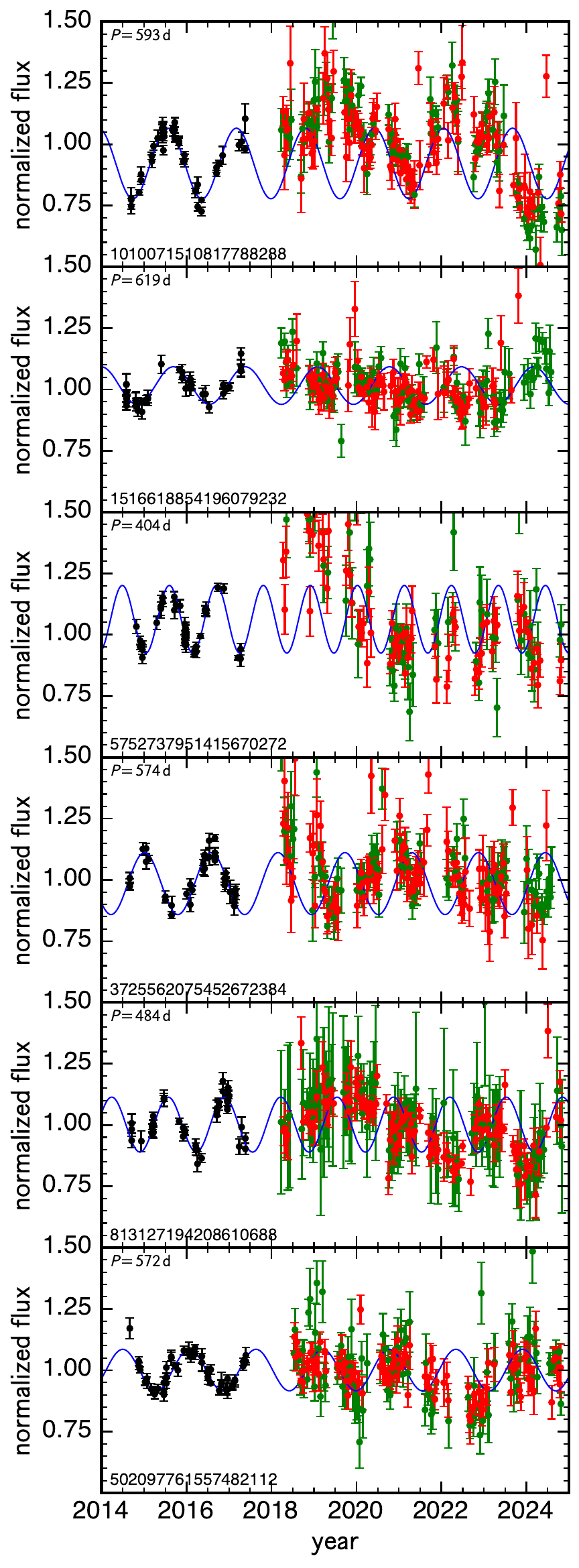}
\caption{Figure 3 (continued)}
\end{figure}

\begin{figure}[H]
\figurenum{3}
\centering
\includegraphics[width=\columnwidth]{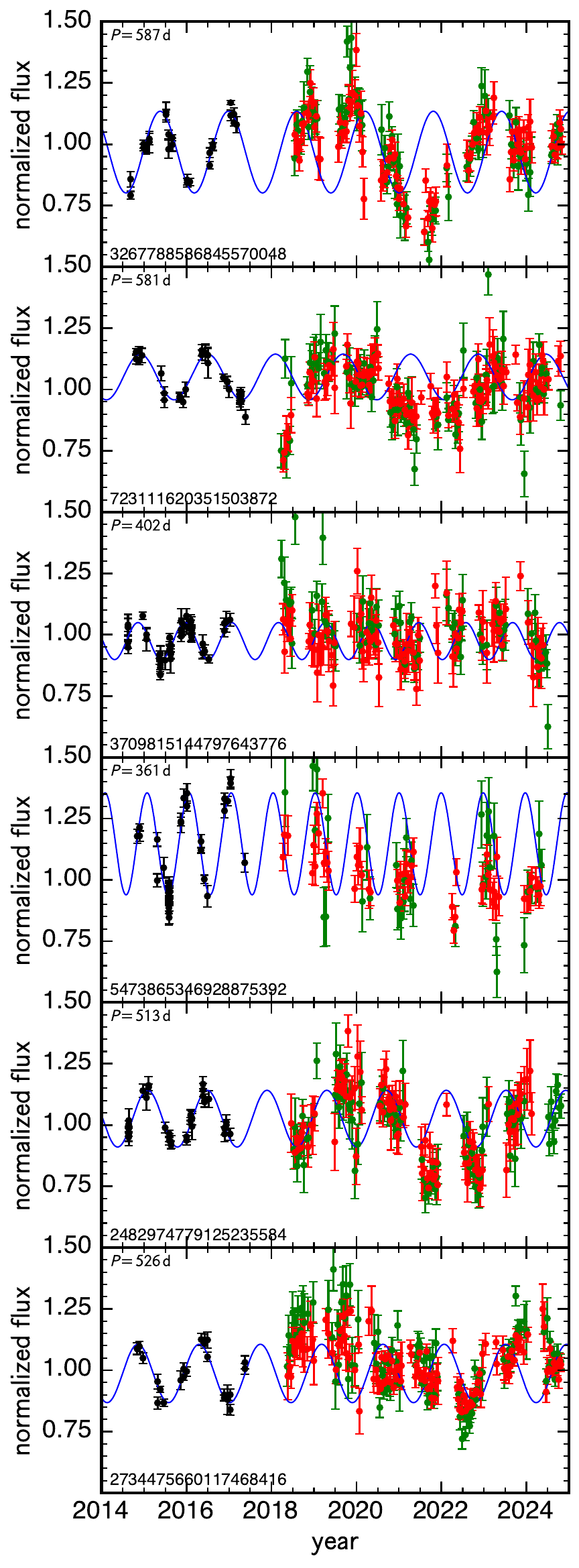}
\caption{Figure 3 (continued)}
\end{figure}

\begin{figure}[H]
\figurenum{3}
\centering
\includegraphics[width=\columnwidth]{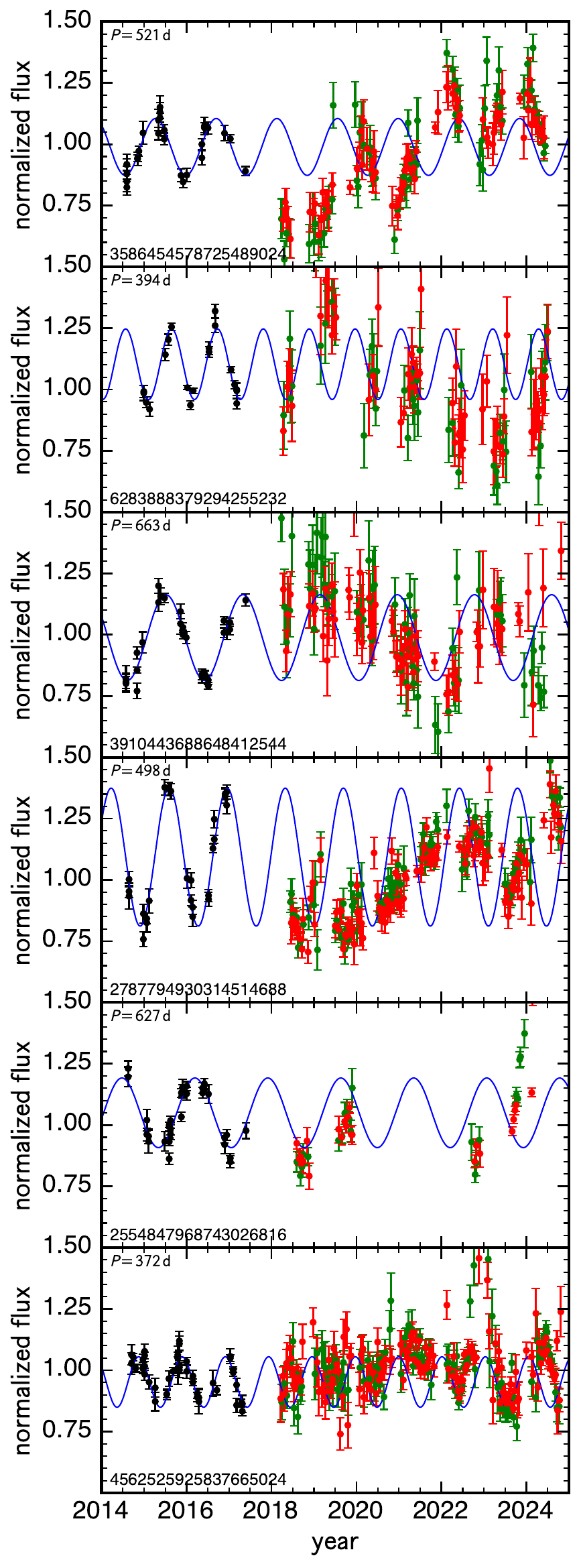}
\caption{Figure 3 (continued)}
\end{figure}

\begin{figure}[H]
\figurenum{3}
\centering
\includegraphics[width=\columnwidth]{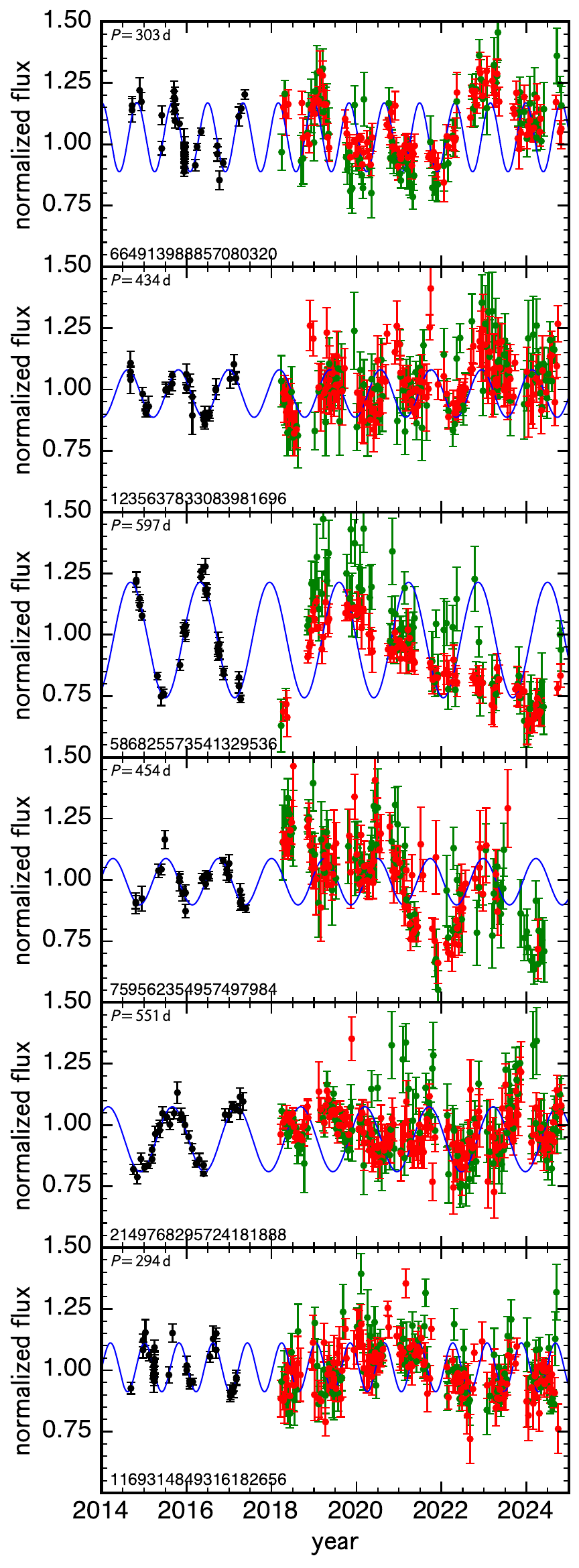}
\caption{Figure 3 (continued)}
\end{figure}

\begin{figure}[H]
\figurenum{3}
\centering
\includegraphics[width=\columnwidth]{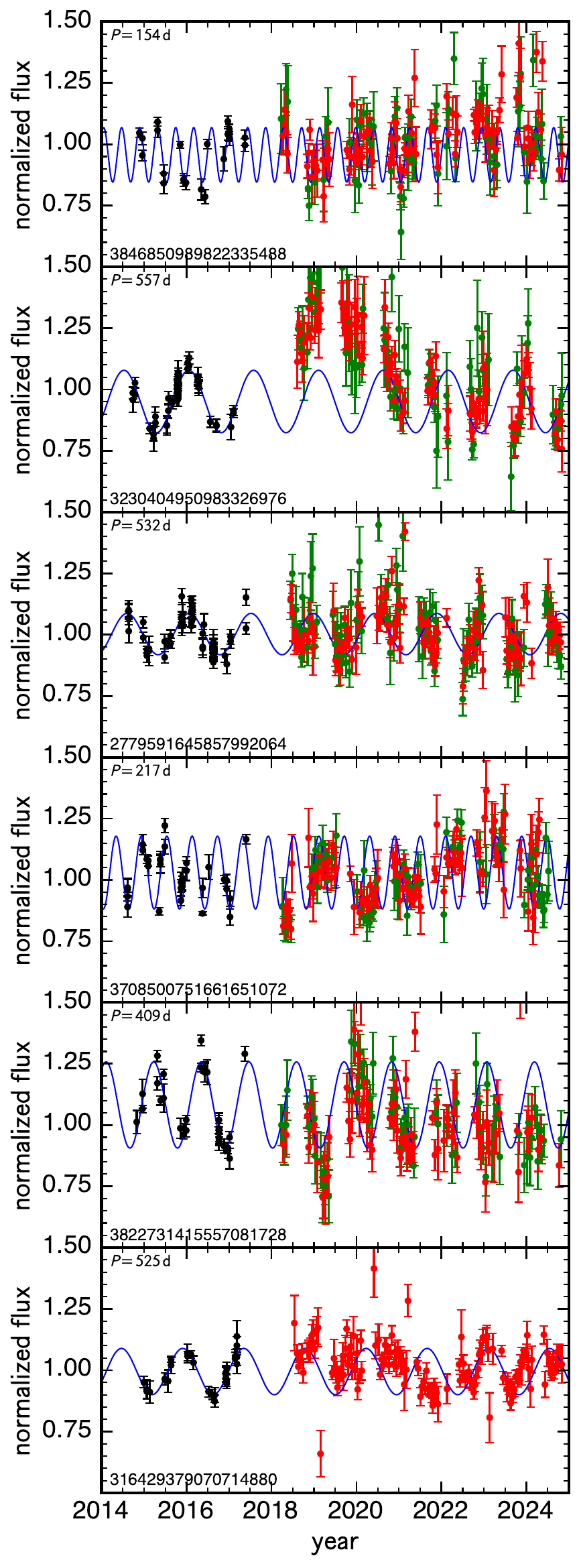}
\caption{Figure 3 (continued)}
\end{figure}

\begin{figure}[H]
\figurenum{3}
\centering
\includegraphics[width=\columnwidth]{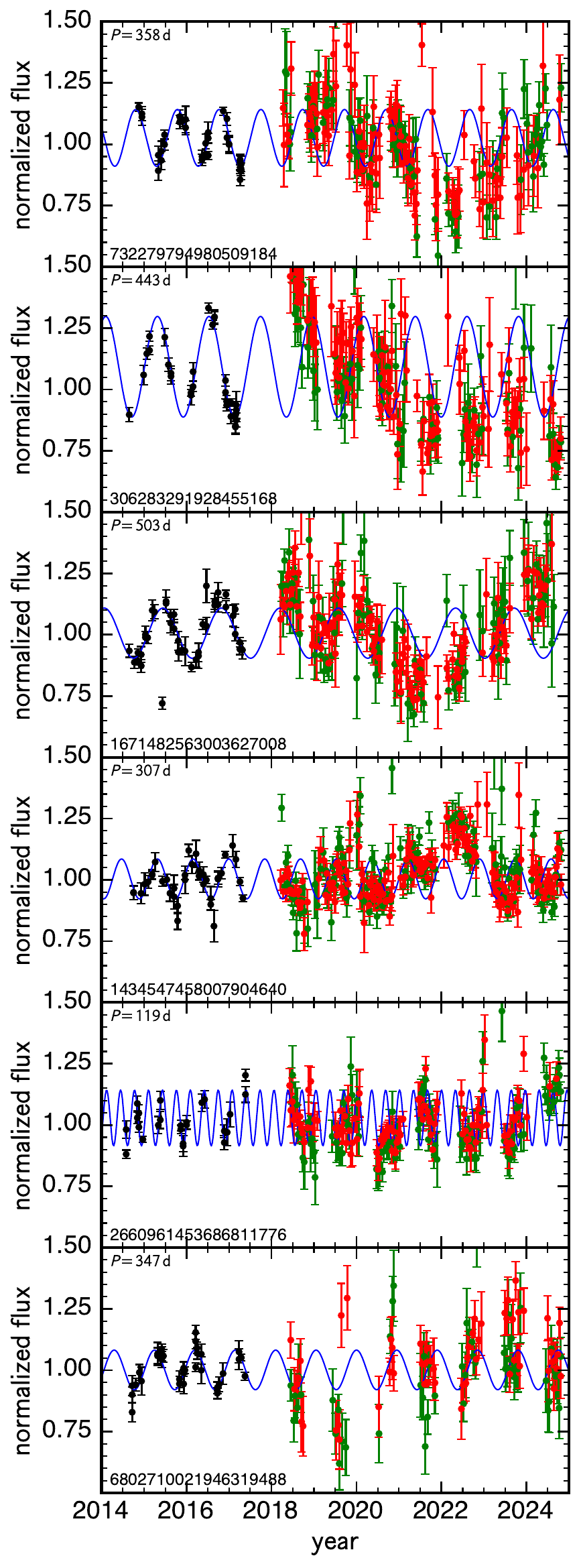}
\caption{Figure 3 (continued)}
\end{figure}

\begin{figure}[H]
\figurenum{3}
\centering
\includegraphics[width=\columnwidth]{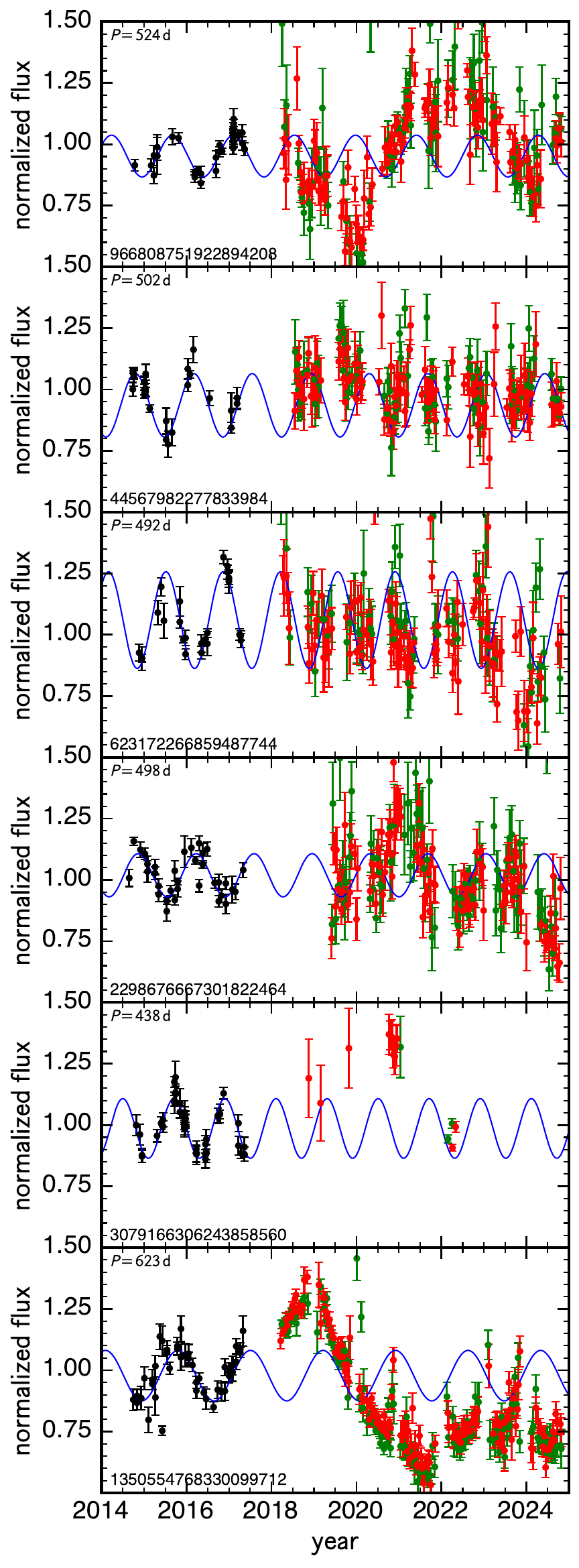}
\caption{Figure 3 (continued)}
\end{figure}

\begin{figure}[H]
\figurenum{3}
\centering
\includegraphics[width=\columnwidth]{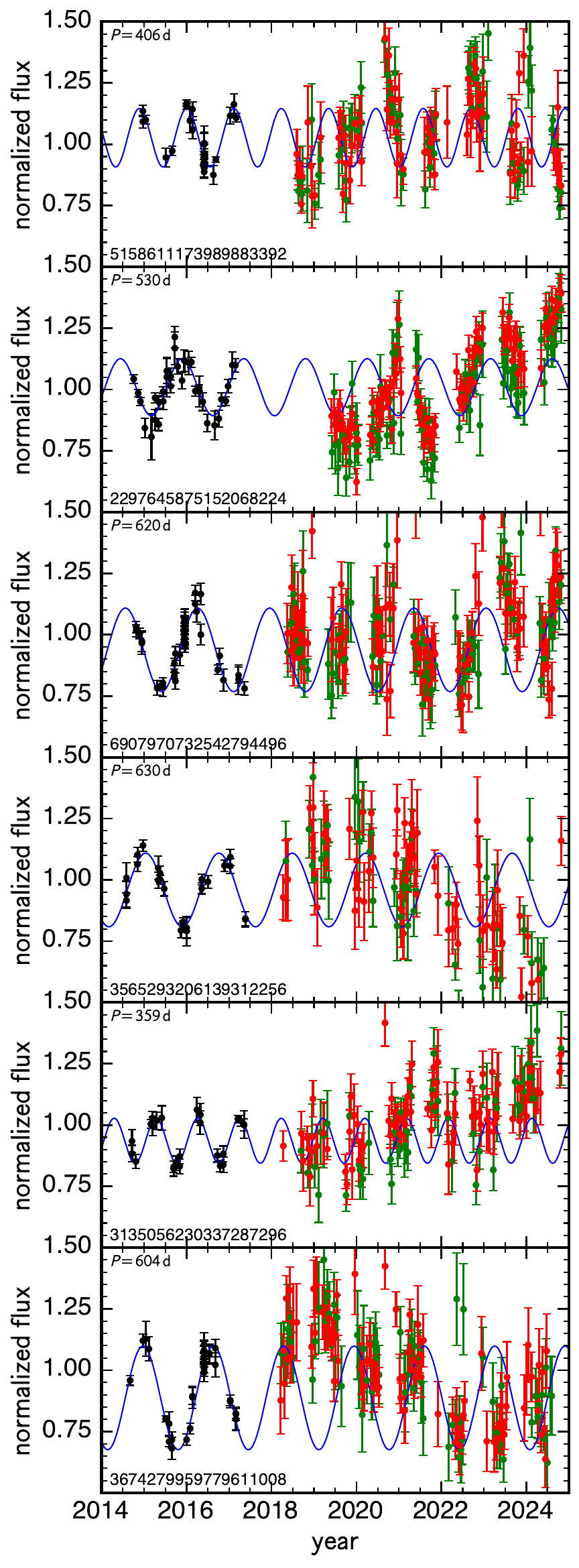}
\caption{Figure 3 (continued)}
\end{figure}

\begin{figure}[H]
\figurenum{3}
\centering
\includegraphics[width=\columnwidth]{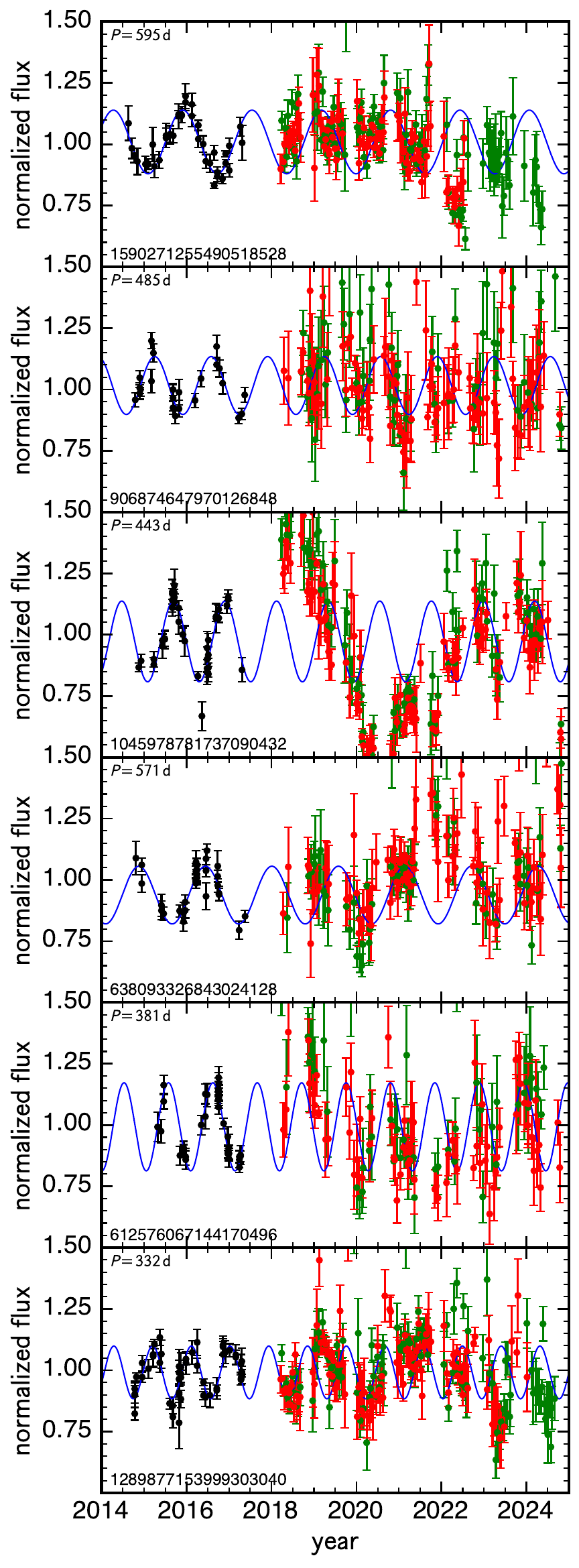}
\caption{Figure 3 (continued)}
\end{figure}

\begin{figure}[H]
\figurenum{3}
\centering
\includegraphics[width=\columnwidth]{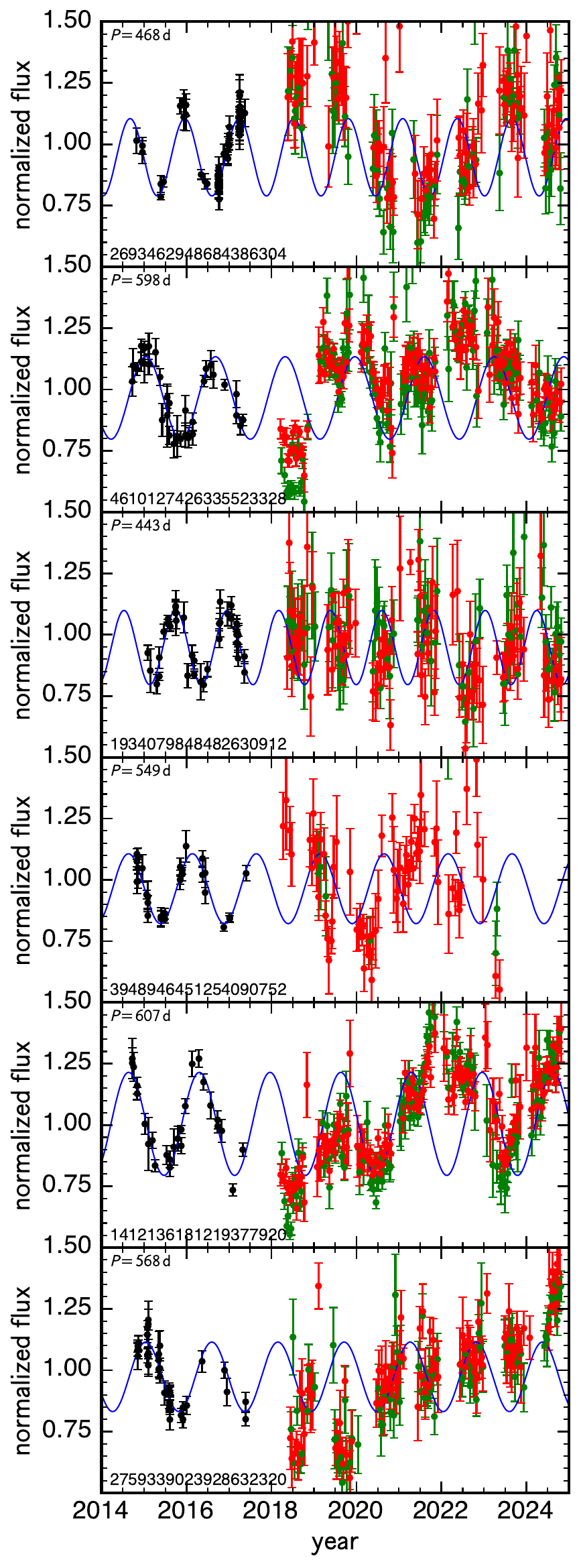}
\caption{Figure 3 (continued)}
\end{figure}

\begin{figure}[H]
\figurenum{3}
\centering
\includegraphics[width=\columnwidth]{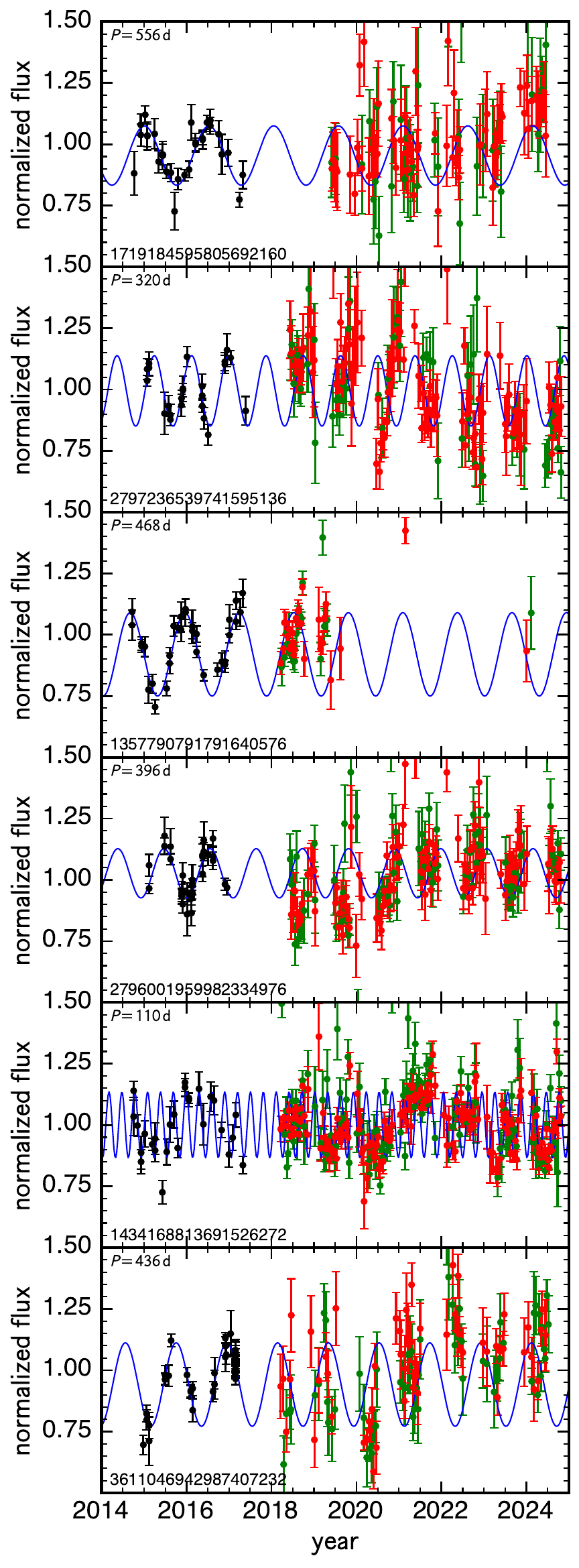}
\caption{Figure 3 (continued)}
\end{figure}

\begin{figure}[H]
\figurenum{3}
\centering
\includegraphics[width=\columnwidth]{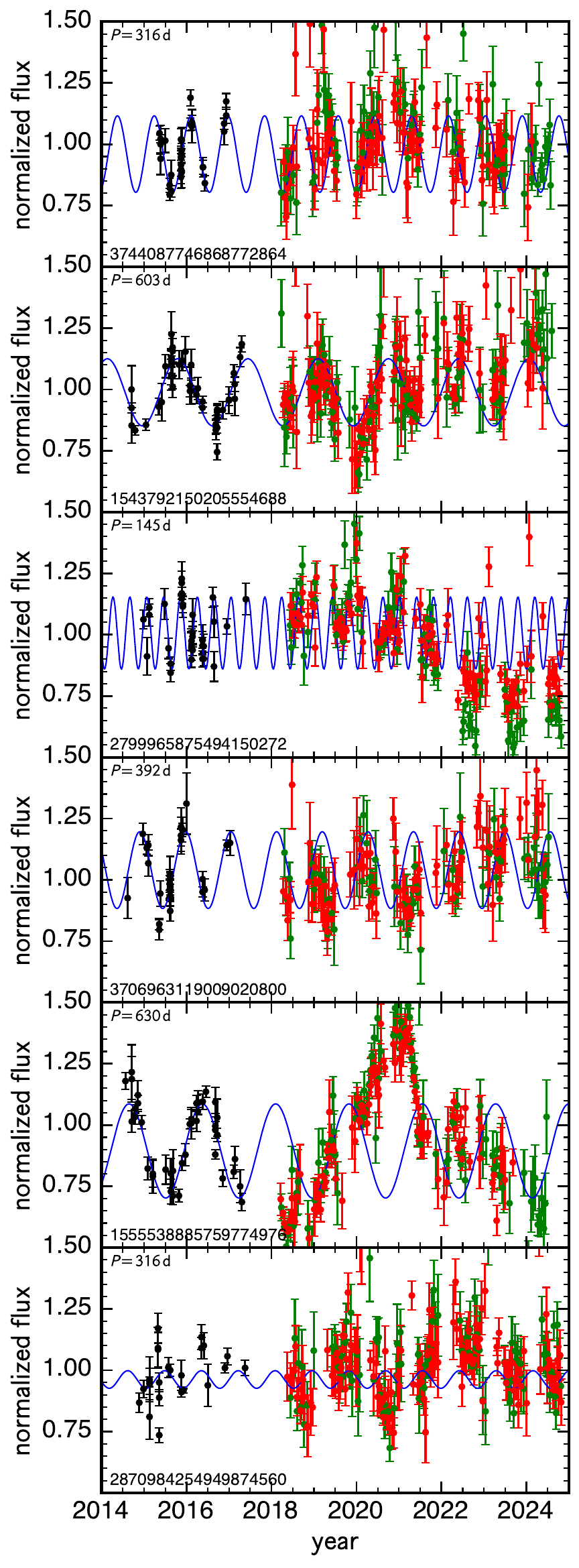}
\caption{Figure 3 (continued)}
\end{figure}

\begin{figure}[H]
\figurenum{3}
\centering
\includegraphics[width=\columnwidth]{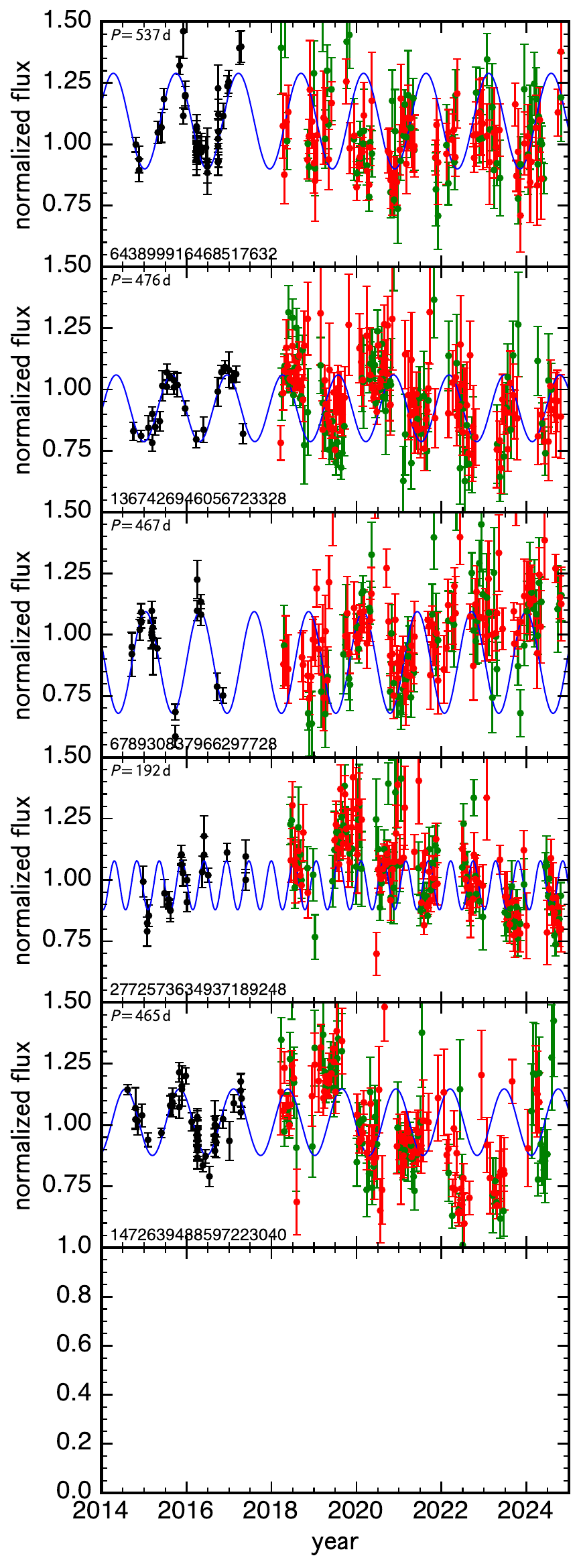}
\caption{Figure 3 (continued)}
\end{figure}

\end{document}